\documentclass[final,5p,times,twocolumn]{elsarticle} 

\usepackage{amssymb}
\usepackage{lineno}

\journal{Nuclear Instruments and Methods}

\newcommand{\Ch}{Cherenkov }
\newcommand{\Gra}{$\gamma$-ray astronomy }
\newcommand{\Grs}{$\gamma$-rays }
\newcommand{\Gr}{$\gamma$-ray }
\newcommand{\fov}{field-of-view }
\newcommand{\bw}{$\Delta\nu$ }
\newcommand{\enu}{$\xi_{\nu}$ }
\newcommand{\bwl}{bandwidth limited pulse }

\begin{document}

\begin{frontmatter}



\title{Detection of elusive Radio and Optical emission from Cosmic-ray showers in the 1960s.}


\author{David J. Fegan}

\address{School of Physics\\ University College Dublin, Dublin 4, Ireland}

\begin{abstract}
During the 1960s, a small but vibrant community of cosmic ray physicists, pioneered novel optical methods of detecting extensive air showers (EAS) in the Earth's atmosphere with the prime objective of searching for point sources of energetic cosmic $\gamma$-rays. Throughout  that decade, progress was extremely slow. Attempts to use the emission of optical \Ch \cite{CH86} radiation from showers as a basis for TeV gamma-ray astronomy proved difficult and problematical,  given the rather primitive light-collecting systems in use at the time, coupled with a practical inability to reject the overwhelming background arising from hadronic showers. Simultaneously, a number of groups experimented with passive detection of radio emission from EAS as a possible cheap, simple, stand-alone method to detect and characterise showers of energy greater than $10^{16}$ eV. By the end of the decade, it was shown that the radio emission was quite highly beamed and hence the effective collection area for detection of high energy showers was quite limited, diminishing the effectiveness of the radio signature as a stand-alone shower detection channel. By the early 1970s much of the early optimism for both the optical and radio techniques was beginning to dissipate, greatly reducing research activity. However, following a long hiatus both avenues were in time revived, the optical in the early 1980s and the radio in the early 2000s. With the advent of digital logic hardware, powerful low-cost computing, the ability to perform Monte Carlo simulations and above all, greatly improved funding, rapid progress became possible. In time this work proved  to be fundamental  to both High Energy \Gr Astronomy and Neutrino Astrophysics. Here, that first decade of experimental investigation in both fields is reviewed.
\end{abstract}

\begin{keyword}

high energy particles \sep gamma rays \sep radio and optical
\end{keyword}

\end{frontmatter}


\section{Introduction and perspective}\label{Intro}
Let me begin by expressing my gratitude to the organisers for inviting me to deliver the inaugural lecture at ARENA (Acoustic and Radio EeV Neutrino detection Activities) 2010.  As the title  suggests, I will review some historical and pioneering aspects of research conducted in the 1960s, focussed on detection of both optical \Ch radiation and radio emission from cosmic ray induced Extensive Air Showers (EAS) in the earth's atmosphere. I had the good fortune to undertake my PhD at University College Dublin (UCD) in the group of Neil Porter, where research on both topics was simultaneously underway. This paper is somewhat personal, based on recollections of a variety of pioneering but challenging experiments,  some of which were successful, others not. The formative experience of working with a number of outstanding cosmic-ray experimental physicists was a unique privilege. Many of the early scientific accomplishments of what later became known as High Energy Astrophysics were achieved by the cosmic ray research community which had its origins in the immediate aftermath of WW II and which had among its numbers, researchers with extensive wartime research experience in the techniques of experimental nuclear physics, fast electronics/signal processing, radar development etc. 

In 1960, Kenneth Greisen \cite{KG60} issued a challenge based on the communitiesÕ obsession with establishing the precise form of the primary cosmic ray energy spectrum and with finding possible origins of cosmic radiation :-
`` If knowledge of the spectrum is to be extended much further ($ \geq 10^{16} $ eV) it will be necessary to adopt new methods of detection based on a type of radiation that permits the showers to be observed over a much wider radius than that reached by charged particles. Use of Cherenkov light offers some improvement but a more isotropic electromagnetic radiation in the visible or radio spectrum will ultimately provide the greatest area of reception''. This prescient observation (together with some earlier work) became a powerful motivating factor in focussing members of the cosmic ray community on examination of  new optical and radio experimental techniques to detect extensive air showers (EAS) initiated by high energy primary cosmic rays. 

It is also important to emphasise the fact that scientific research funding in Ireland during the 1960s was very scarce, with researchers dependent on sporadic, minuscule, grace-and-favour funding from within the university system, as no formal national funding agency existed at that time. Fortunately for much of that decade, Neil Porter was in possession of a modest USAF research grant, administered from the European Office of Scientific Research. This grant helped in purchasing essential specialist items such as oscilloscopes, photomultipliers, communications equipment and also facilitated limited travel support by the group. Nevertheless, given the austerity of the era, highly imaginative and innovative low-cost solutions had to be found in order to solve many of the technological problems that regularly surfaced. Much of the signal processing electronics, the particle detectors, radio antennae and hardware infrastructure, had to be constructed by a combination of graduate students and machine shop technical staff. In terms of electronics, this was very much an era of transition, from power consuming vacuum tubes to solid state devices such as transistors and diodes. It was only towards the close of the 1960's, that digital logic chips and primitive hybrid analog linear chips became available, often at very considerable cost. This meant that much of the detector and electronic system development had to be designed and assembled using discrete components manually fabricated onto copper-strip circuit boards.  Building a 2ns risetime analogue pulse amplifier was not for the feint-of-heart! There were no data-acquisition systems, no online computers, no DAC, ADC or DSP chips. However, our group benefitted hugely throughout that decade, from a formal collaboration with the group of John Jelley based at AERE (Atomic Energy Research Establishment), Harwell, UK, which provided direct access to the transistor based Harwell 2000 series modular electronic family. 
\section{Early \Ch detector systems 1960-66} \label{S:earlyC}
\subsection{First photomultiplier based and image intensifier based detectors} \label{SS:FirstPM}
\begin{figure}
\begin{center}
\includegraphics[width=.5\textwidth]{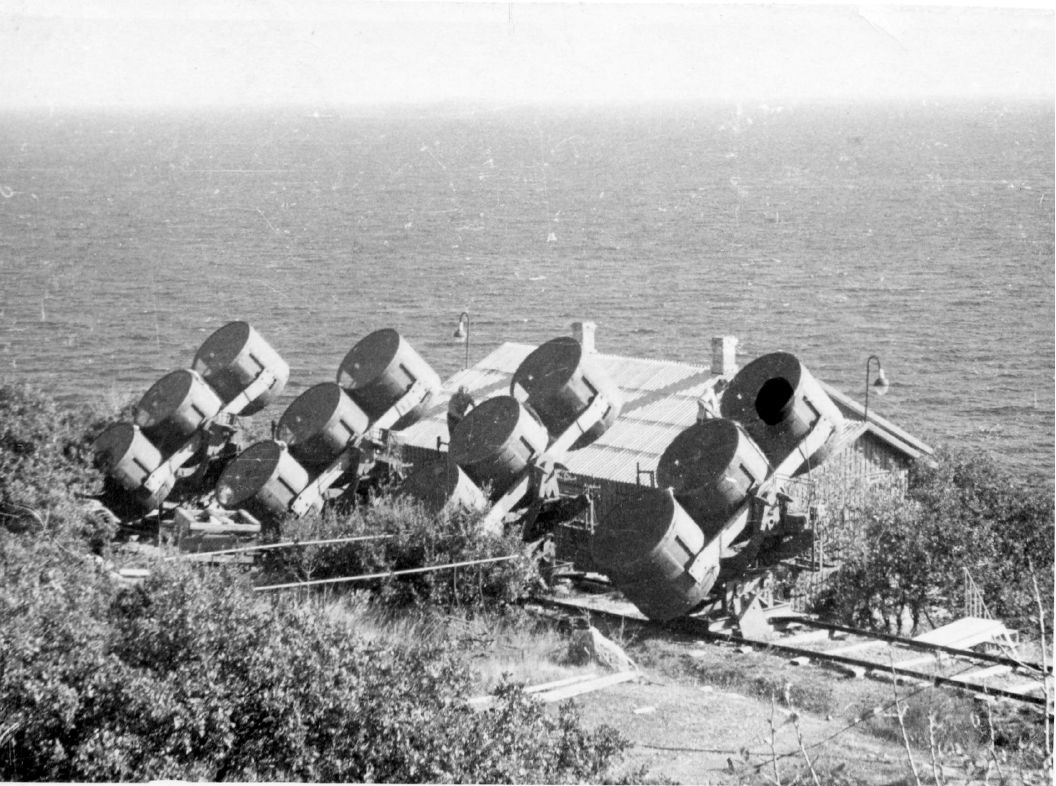}
\caption {Crimean \Ch installation of A.E.Chudakov, on the shores of the Black Sea, early 1960s.}
\label{fig:Crim}
\end{center}
\end{figure}

John Jelley was a hugely influential and respected physicist, who, together with his colleague Bill Galbraith,  first successfully detected \Ch pulses from the night sky using a WW II signalling mirror, viewed by a photomultiplier tube and  mounted within a dustbin. They observed bandwidth-limited pulses on a free-running, un-triggered oscilloscope timebase, at a rate of about 1 to 2 per minute \cite{GJ53}. Being a fastidious, meticulous and most careful experimentalist, John went on to study various aspects of \Ch radiation in both liquids, solids and gases before writing his classic text book `` \Ch Radiation  \emph{and its applications} ''  \cite{JVJ58} in 1958. In time, this book became compulsory reading material for generations of graduate students in various fields. The pioneering \Ch experiments are described in some detail in Jelley's 1982 paper \emph{``Flashes In a Dustbin and Other Reflections"} \cite {JVJ82}. The findings of Galbraith and Jelley were independently confirmed within the USSR in 1955 \cite{NC55}. The 1958 paper of P.Morrison on X-ray and $\gamma$ -ray astronomy \cite{M58} together with the prediction of Cocconi \cite{Coc}  of a possible strong flux of \Grs from the Crab Nebula, were important stimuli to both emerging research fields.  It was clear to A.E.Chudakov and collaborators that the \Ch technique was a potentially powerful tool for  \Gra in the energy range $10^{11}$ to $10^{12}$  eV. A remarkable experiment conducted in the Crimea over four winters (1960-63) by the cosmic-ray group of the Lebedev Physical Institute utilised an array of 12 individual mirrors each of 1.5m diameter and $3.5^\circ$ field-of-view, shown in  Figure  \ref {fig:Crim}. Ten target sources including the Crab Nebula, Cygnus A and Virgo A were observed (unsuccessfully) establishing flux upper limits  $\simeq$  5 x $10^{-11}$ photons cm$^{-2}$  s$^{-1}$  \cite{Ch65}.

\begin{figure}
\begin{center}
\includegraphics[width=.5\textwidth]{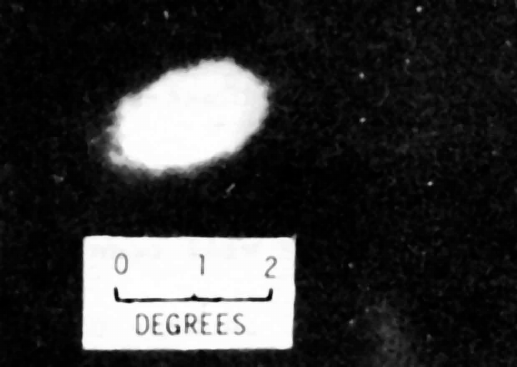}
\caption{Intensifier image of shower.}
\label{fig:Inten}
\end{center}
\end{figure}

Contemporaneously with the Crimean work, Neil Porter's research group at University College Dublin was initially concentrating on photography of \Ch\ light from showers using image intensifiers. The prime scientific objectives were to learn more about shower development in the atmosphere and hopefully, in the process, to exploit large shower collection areas through use of \Ch\  optical detectors. Despite the very difficult technical challenge, a series of remarkable photographs of shower images were recorded at Agassiz during 1961, see Hill \& Porter \cite {HP61} and at Mt. Chacaltaya during 1962, see Hill et al. \cite {Hetal}. The dynamic range of detected shower energies was from $10^{15} $ to $10^{16}$ eV and Figure  \ref {fig:Inten} is representative of one of the more energetic events detected. Larger showers were characteristically comet-shaped, comparing favourably with analytical models based on electron-photon cascades. However, given the very high energy threshold of the technique, coupled with operational difficulties and the overwhelming hadronic shower background, there was little hope or possibility of performing point-source neutral cosmic ray searches and image intensifier applications were suspended around 1963. The pioneering work on intensifier based energetic shower imaging is well captured in Porter's later review article \cite {POR82}. Contemporaneous with the early intensifier measurements, Boley \cite{Boley64} used a single mirror reflector and a matrix of 19 photomultiplier tubes, in order to elucidate the angular spread of \Ch light from EAS. 
\subsection{ The joint Harwell-UCD \Ch system}
The possibility of detecting high energy \Grs from discrete sources was deemed by Jelley and Porter to be of fundamental astrophysical interest, since observations of optical and radio synchrotron emission do not establish particle energies but simply a combination of energy and magnetic field. They realised that observations of \Grs above $10^{11}$ eV would be of greater significance than measurements in the rocket and satellite regime $\simeq 10^{9}$ eV; establishing both the magnetic field and particle energies more precisely. So, even the establishment of upper limits for \Grs emission would have scientific merit by ruling out possible mechanisms through which high energy electrons are generated. Such considerations led to the establishment in 1961, of an informal collaboration between both groups. The discovery of quasars in 1962, coupled with the growing interest in high energy particles associated with powerful radio sources, simply accelerated efforts to improve the efficacy of the \Ch technique in the energy range $10^{11}$ to $10^{13}$  eV and a formal collaboration between the Harwell and UCD groups was entered into in 1963, committing both groups to development of photomultiplier-based systems while simultaneously abandoning intensifier development.

Almost immediately, a most influential paper was published by Jelley and Porter -  \emph {``\Ch Radiation from the Night Sky, and its Application to $\gamma$-Ray Astronomy"} \cite{JP63}. In this paper many fundamental technical  issues were discussed including, (i) the possibility of detecting \Gr primary particles through detailed examination of the structure of the shower, since as the shower axis falls at increasing distances from the optical detector's axis, there is a progressive displacement of the peak intensity from the true shower direction. At 100m, it is about 0.7 $^\circ$ for a \Gr shower and 1.3 $^\circ$ for a nuclear shower, (ii) evidence from intensifier pictures of higher energy showers indicating the importance of the shower spot \emph{shapes} in supplying useful information on true shower directions,  (iii) although the angular resolution and the detector \fov are one and the same thing, thus restricting operation to small patches of sky at any given pointing of an instrument, nevertheless clustering of groups of photomultiplier tubes at a detector focus might possibly ameliorate this inherent limitation, thereby improving angular resolution, and (iv) in the context of discriminating \Gr point sources from the overwhelming cosmic-ray background, the real benefit of operating a pair of separated \Ch detectors in \emph{stereoscopic} viewing mode.

\begin{figure}
\begin{center}
\includegraphics[width=.5\textwidth]{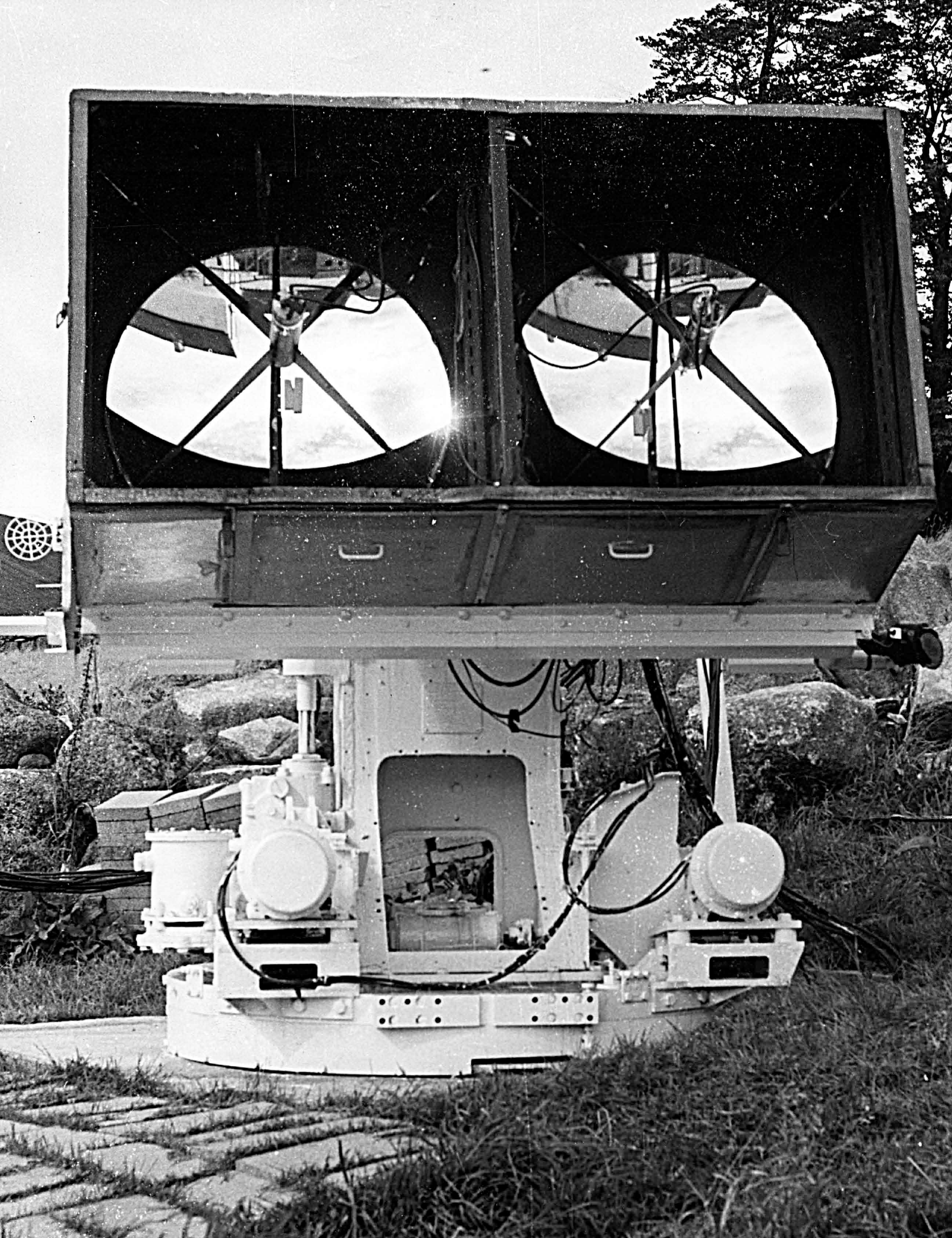}
\caption {UCD-Harwell 2-fold \Ch system at Glencullen in the Dublin mountains (1963-1967).}
\label{fig:DET63}
\end{center}
\end{figure}

\begin{figure}
\begin{center}
\includegraphics[width=.5\textwidth]{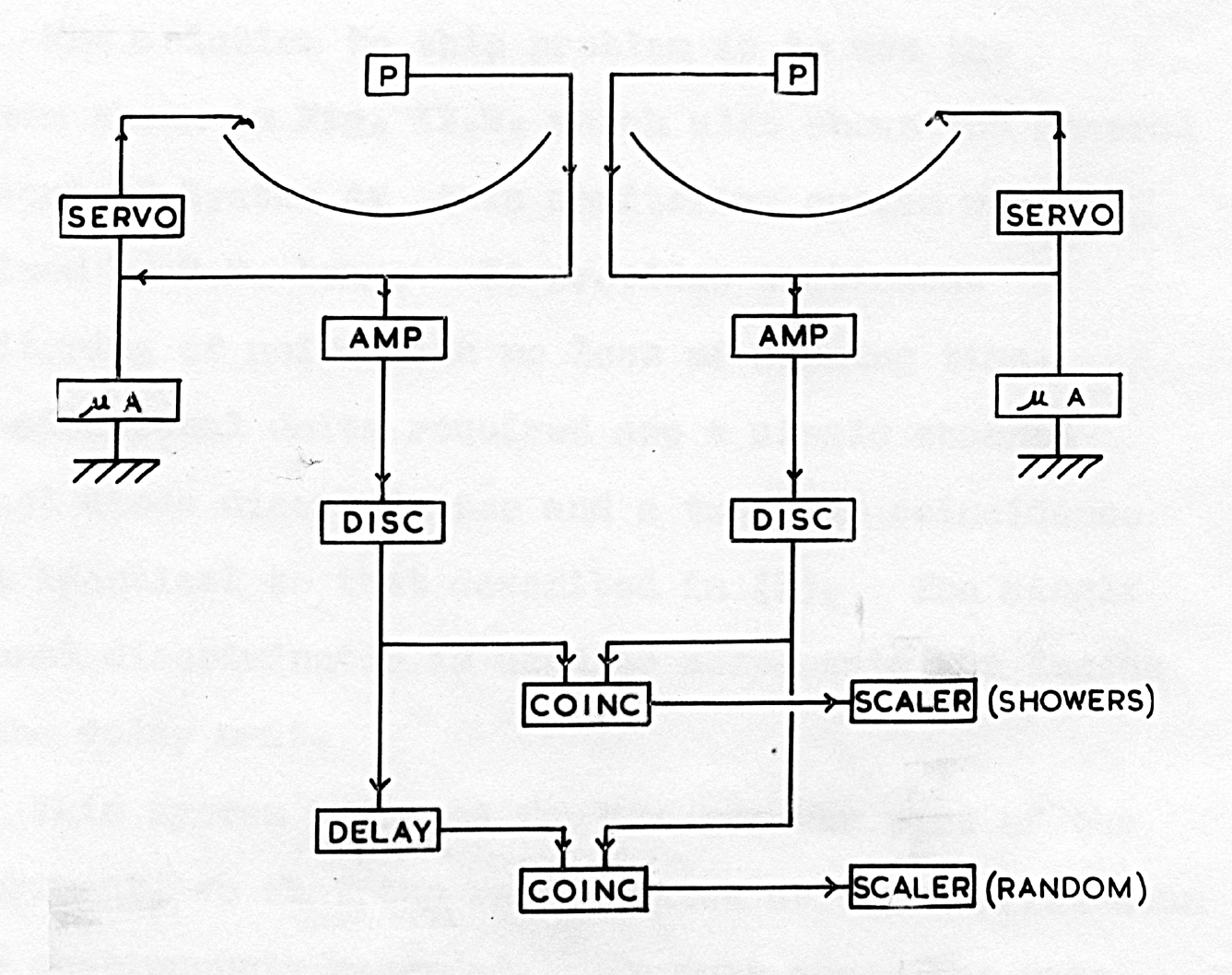}
\caption{UCD-Harwell signal processing electronics.}
\label{fig:DETELECT63}
\end{center}
\end{figure}

In terms of a practical \Ch detector system design, it was considered desirable to combine the highest possible flux sensitivity with the lowest possible energy threshold. Jelley  \cite{JVJ58} had earlier shown that in the absence of electronic pile-up, the threshold energy of a typical single mirror system is  $ E_t  \propto D^{-1} \; \Omega \; \Delta \nu^{-1}$ where D is the optical receiver diameter,  $\Omega$ is the \fov and \bw is the amplifier bandwidth. A prototype detector was developed at Harwell during 1962-63 and relocated to the sheltered rural valley of Glencullen, Co. Dublin in 1963. The mirror mount consisted of a WW II Bofors gun mounting (3.5 tons) which supported twin 90 cm. diameter back-silvered mirrors from searchlights, adequate optically for use with the fastest photomultiplier tubes then available (2 inch RCA 6542). The mounting and mirrors are shown in Figure \ref{fig:DET63} while the primitive signal processing electronic system is depicted in Figure \ref{fig:DETELECT63}.


Occasionally this system was operated in ON-OFF mode by pointing ON a candidate source and allowing the source to drift through the \fov, typically 12 to 15 minutes duration. Then the system was redirected in azimuth, OFF the source, and a comparison \emph{null}  observations taken.
However, this system was mostly operated in drift-scan mode whereby the mounting was set well  ahead of the source,  allowing the earth's rotation to carry target sources through the detector's \fov and a symmetric distance beyond the source. Twin servo systems stabilised the individual photomultiplier currents to constant values, through addition of about 20\% additional light, using small incandescent filament lights mounted close to the photomultipliers. This expedient helped compensate inherent variations in sky brightness across source regions, as drift-scans progressed. The data capture was remarkably primitive. The operator simply recorded both \emph{prompt} (genuine coincident showers plus random background showers) and \emph{random} (delayed or random coincidences, representing background) events, on a minute-by-minute basis, through visual inspection of a pair of scalers - the numbers simply being written into the nightly log, by hand. No additional information on individual showers was capable of being captured. Any manifestation of  source activity was simply reflected in the difference between the two count rates as a function of time.

The UCD-Harwell \Ch system had a threshold energy comparable to that of the Crimean system \cite{Ch65} but was considerably smaller in size. The minimum detectable flux of \Grs was estimated as being in the range $10^{-10}$  to $10^{-11}$ photons $\rm cm^{-2} \rm s^{-1}$. The list of target objects selected for study to some extent concentrated on alternative objects not previously viewed by the Lebedev group, lying in the declination band +20 to + 40 and constrained to a right ascension sky strip from 20 hr through 24 hrs to 15hrs. The topicality of the discovery of quasars greatly influenced the final list of 13 objects chosen for viewing. The target sources were: the quasars 3C9, 3C48, 3C196, 3C147, 3C273, 3C286; the Crab Nebula, magnetic variables HD71866 and 53 Cam; and the galaxies Cygnus A (difficult location), M31, NGC4472 and 1607+26.  Quasars, because of their high optical continuum emission, generated considerable optimism within the group, given the intrinsic source energetics as estimated by radio and optical astronomers and by theorists. Synchrotron or inverse-Compton emission mechanisms were invoked to explain the continuum emission, hence high-energy electrons were implied in these sources, auguring well for detection of possible TeV \Gr emission.

It just cannot be overstated as to how demanding this experiment was from the operational perspective. While the site offered superb viewing on cloudless, moonless nights, the influence of nearby hills and mountains did generate many cloudy nights where operation was compromised or impossible. During the three successive winters of 1963, 1964 and 1965, only 75 useful fully-operational nights were obtained, with an additional 57 abandoned or utilised for technical purposes. 

The scientific accomplishment of the three seasons observations was quite disappointing \cite{Long65}. In Table  \ref{Tb:Stats1}, the gross findings are summarised, observations being classified in three categories, quasars, non-quasars and overall totals. Observations of individual sources tended to be a mixture of both ON/OFF and drift scans. The quasar results when summed, reflected a slight positive effect. No individual source was observed as being statistically significant. Individual source flux upper limits of the order of $10^{-11}$ photons cm$^{-2}$  s$^{-1}$ were established at threshold energies of the order of  $10^{13}$ eV. The limit on the Crab Nebula was judged to exclude the possibility of a nuclear origin for any  high energy electrons present at these energies.  An important conclusion was that since most of the theoretical mechanisms capable of explaining emission of \Grs from astrophysical objects implied steep energy spectra, it would be essential for any follow-up detector systems to bring down the detection energy thresholds of future detectors. Interestingly, during these years, the possibility of using passive radio detection techniques as a tool in cosmic ray studies became an exciting distraction for the Harwell-UCD collaboration, from the disappointments of scientific progress with the optical work.

\begin{table}
\caption{Combined source statistics 1963-64; 1964-65; 1965-66. (a) Counts on sources; (b) Counts off sources;
(c) Total number of observations; (d) Number of observations giving a positive effect;  (R) ON/OFF ratios; (SE) standard errors.}
\label{Tb:Stats1}
\vspace{0.1in}
\begin{center}
\begin{tabular}{|c||rrrr||c|c|} \hline
SOURCE        & a        & b         & c    &d                & R       & SE \\ \hline  \hline
Quasars         & 23195 & 22717 & 87  & 55             & 1.021 & 0.009 \\
Non-Quasars & 9118   & 9308   & 56   & 19             & 0.980 & 0.015 \\ \hline
Totals             & 33313 & 33025 &143  & 74            & 1.009 & 0.008 \\ \hline 
\end{tabular}
\end{center}
\end{table}

\section{Radio Emission from EAS}
\subsection{Discovery experiment at Jodrell Bank 1964-65}

Jelley in his book on Cherenkov radiation \cite{JVJ58} and in a subsequently little known paper published in 1962 \cite{JVJ62}, considered whether the Cherenkov emission mechanism that gives rise to optical emission with a $\nu$\bw spectrum, might also radiate in the microwave region of the spectrum and might possibly be detectable with sensitive receivers. His conclusions were pessimistic, largely due to the phenomenal reduction in both frequency and bandwidth as one moves from optical frequencies down to the microwave regime. Furthermore, when the observational wavelength becomes greater than the typical charge separation in an EAS, then radiation from individual positive and negative charges tends to cancel out and no net electromagnetic field ensues, assuming overall charge neutrality prevails in disk of shower particles as it sweeps down through the atmosphere. \\

Somewhat serendipitously, that same year (1962) saw a paper published by Russian theorist G.A.Askaryan \cite{GA62} who drew attention to the consequences of high energy particles colliding with a dense medium such as rock on the moons surface. His prime objective was to draw attention to the fact that rock is a dielectric material essentially transparent to radio waves and the moon might therefore provide a large target mass for possible detection by radio, of high energy cosmic neutrinos interacting in the rock. But most importantly, Askaryan also pointed out that due to the preferential annihilation of shower positrons ``in flight '' by interaction with electrons in the dielectric medium, a natural negative charge excess would arise in the shower front. The importance of this charge excess from the perspective of terrestrial radio detection of EAS in the earth's atmosphere was not immediately appreciated. In essence, the net negative charge excess, taken in conjunction with Compton recoils and delta rays, effectively constitute a localised group of electrons with energies between 2 and 30 MeV, capable of radiating as a coherent bunch of particles. Assuming N particles populate the shower at maximum, then some charge excess $\epsilon $N builds up.  For incoherent radio emission the signal intensity is proportional to N times that of a singly charged particle but for coherent emission the signal is proportional to ${\epsilon}^2$${N}^2$, so the gain is ${\epsilon}^2$N. If N=$10^6$ and if $\epsilon \simeq 0.1$ then the enhancement factor is $10^4$, clearly an enormous coherent radiation gain factor over the incoherent situation. In order to preserve coherence, the radiating particles all have to be at the same distance from the detecting antenna, to an accuracy of a small fraction of a wavelength. The longitudinal dispersion of the shower particles is effectively the shower disk thickness of between 2m and 3m,  as the shower sweeps to earth. The coherence condition requires therefore that the wavelength of observation be greater than the physical dimensions of the emitting region. At a frequency of 75 MHz the corresponding wavelength is 4m, so the observational wavelength needs to be greater than this value for detection of a coherent signal. 
\begin{figure}
\begin{center}
\includegraphics[width=.5\textwidth]{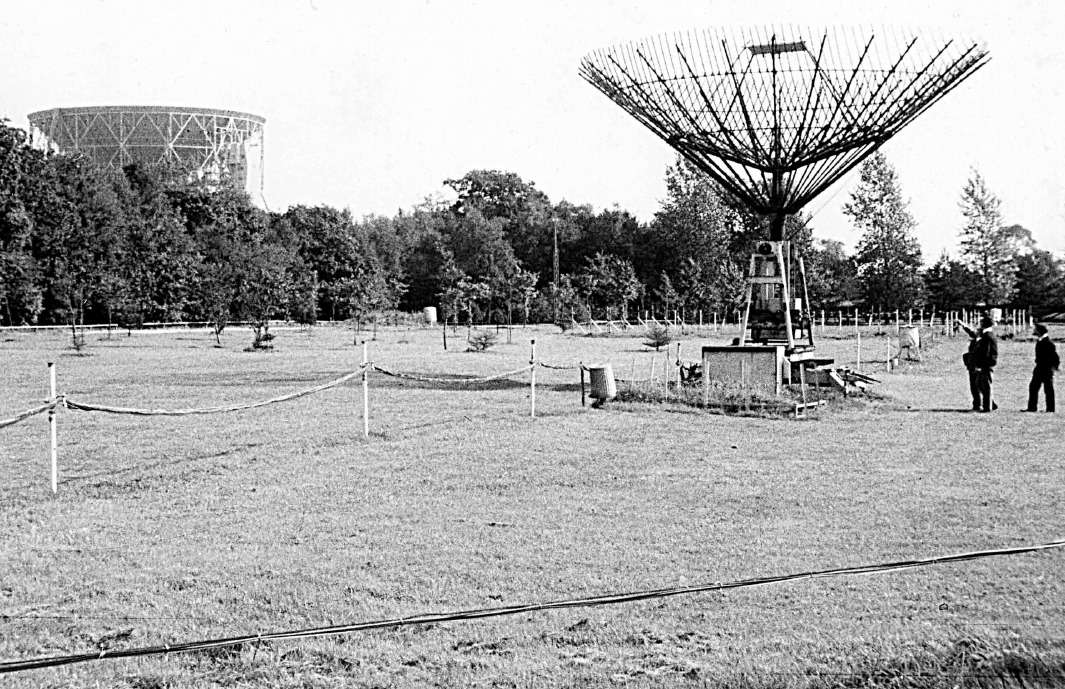}
\caption{Jodrell Bank dipole array and particle detectors, photographed in early 1967.}
\label{fig:JODRELL}
\end{center}
\end{figure}
The elegant and successful prototype experiment that led to the discovery of radio pulses from EAS \cite {JVJ65} and \cite {JVJ66} was conducted by a collaboration between research groups from Harwell, University College Dublin and University of Manchester, at Jodrell Bank in 1964, (Figure \ref{fig:JODRELL}), with the historical details described in considerable detail by Weekes  \cite{TCW2001}. Using a broadside array of 72 dipoles (44 MHz, detector bandwidth 2.75 MHz)  directed at the zenith and triggered by a small Geiger counter array, 11 unambiguous radio pulses were detected from about 4,500 particle array triggers as oscilloscope timebase traces, establishing the veracity of the effect. The shower array trigger threshold was $5 \times 10^{16}$ eV and the typical energy of individual radio pulses was only 1eV implying a total radio energy content from showers of energy close to threshold, of the order of 1 part in $10^8$  of the total shower primary energy. Soon afterwards an elucidation of possible mechanisms responsible for radio emission was made by Kahn and Lerche \cite {KL66} whereby three distinct physical processes at work in the shower were discussed Ð
(a)	radiation from the net charge excess of shower electrons;
(b)	a geomagnetically induced electric dipole moment arising from Lorenz separation of positive and negative charge;
(c)	a geomagnetically induced transverse current within the disk.
So, on the basis of a somewhat idealised shower model the findings of Kahn and Lerche suggested that the dominant physical process is the transverse current flow, for nearly all observational frequencies and distances from the shower core. The charge excess process is about an order of magnitude less efficient, while the dipole mechanism becomes important at higher frequencies and for observations made close to the core.
\subsection{Radio: New probe for EAS detection, or not?}
Following the successful Jodrell experiment and an independent conformation at 70 MHz by the UCD group \cite{POR65} at the Glencullen site in the Dublin mountains, the apparent efficacy of the method immediately prompted new participants to enter the field, primarily on the basis of modest start-up costs and simplicity of the experimental infrastructure. The driving scientific motivation was the prospect of using passive radio techniques in order to detect very energetic showers and to measure the size spectrum of showers, with a secondary objective of determining the primary mass composition of showers. One one hand, while the phenomenological aspects of the radio emission required careful study, on the other the tantalising prospect of shower detection by radio techniques alone was a powerful motivating factor. However, the presence of unwanted interference from man-made terrestrial sources was of concern and it was appreciated that the background problem might vitiate against \emph{ever} simply using radio receivers alone, for shower detection. Fast timing, real-time digital signal processing and digital data-logging were all technologically out of the question at that time. 

Fairly quickly it became evident that if radio was to be a useful supplemental tool for cosmic ray research then radio detectors needed to be backed up with ancillary cosmic ray detection equipment such as Geiger counter arrays, scintillator shower arrays and/or optical Cherenkov systems. Two experimental philosophies emerged. Existing large air shower arrays such as BASJE (Bolivian Air Shower Joint Array - University of Michigan), Moscow State University and Havarah Park (University of Leeds) were early and obvious choices for locating radio receivers. Large EAS arrays measure many shower parameters including local shower particle densities in multiple detectors, shower zenith angle, core location and impact point on the ground, primary energy etc. In this way a comprehensive picture of each shower could be built up and correlation with radio pulse amplitudes and frequencies could slowly be established. Fluctuations in radio signal characteristics on a shower-by-shower basis (at a given energy and frequency) could only be pursued through having access to array data and measured shower parameters. At the same time, groups lacking access to full-blown arrays, tended where possible, to concentrate on particular specialised aspects of the radio emission process such as low-frequency emission characteristics at 2-3MHz, 22 MHz (University of Calgary), polarisation studies and frequency spectrum (University of Manchester), ultra high frequency emission (University College Dublin, Harwell and University of Manchester)  and frequency spectrum (University of Bologna).  It is neither feasible or fruitful to summarise here the experimental findings of all these radio orientated groups. A most comprehensive review article by Harold Allan was published in 1971 \cite{Allan71} and is highly recommended. However, in keeping with the spirit of this paper reviewing both optical and radio work conducted in the 1960s, two individual strands of radio-orientated research will be described in a some detail, ostensibly with a view to capturing the spirit and optimism of the pioneering work done during the latter half of the 1960s. 
\subsection{Geomagnetic charge separation, radio polarization and radial dependence studies at Haverah Park } 
\begin{figure}
\begin{center}
\includegraphics[width=.35\textwidth]{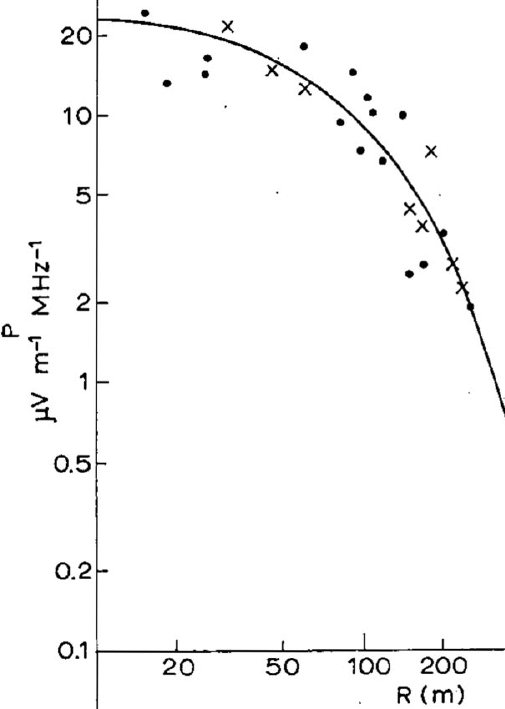}
\caption{The functional dependence of the radio pulse amplitude P on distance R (see text). The ordinate gives normalised field strength for a $10^{17}$ eV shower travelling $\perp$ to the earth's magnetic field. $\bullet$ shows showers where primary energies are in the range $10^{17} \leq E_P \leq 3 \times 10^{17}$ eV, while   x represents the range $3 \times 10^{17} \leq E_P \leq 10^{18}$ eV.}
\label{fig:HAVERAH-1}
\end{center}
\end{figure}

The power of an established EAS array is exemplified by the capability of the Haverah Park array \cite{Tenn68} to establish shower parameters for correlation with radio studies conducted by a collaboration between members of the University of Leeds and Harold Allan of Imperial College London, \cite{Allan71}  \cite{Allan70}.  Radio systems were operated at 32, 44 and 55 MHz with twin orthogonally mounted (N-S and E-W) pairs of antennae operating at each frequency. In order to minimise background noise, the quieter 55 MHz systems were used as selectors, in the interrogation of the outputs of the 32 and 44 MHz receivers. The scientific objective was to establish the geomagnetic origin of the radio emission, whereby the detected pulse for any given shower is maximum when the shower axis is orthogonal to the local component of the earths magnetic field \overrightarrow{B}. About 100 showers were selected in the energy range $10^{17} \leq E_p \leq 10^{18}$ eV and with shower axis impacts R in the range $ 30 \leq R \leq 300$ m from the antennae. For each shower, the array measured the zenith angle $\theta$ and  $sin(\alpha)$, where $\alpha$ is the angle between the shower axis and \overrightarrow{B}. Radio pulse amplitudes \enu  were recorded on film for all radio receiver channels. The following facts were established, (i) polarization measurements were absoluetly consistent with a geomagnetic charge separation mechansim for radio emission; (ii) radio pulse amplitudes \enu  $ \propto sin(\alpha)$; (iii) For $\theta \leq 35^\circ$ and for a limited range of R, \enu $ \propto E_P$, the shower primary energy; (iv) The normalization ratio P = $ \frac{\xi_{\nu}}{E_Psin(\alpha)}$ exhibited the functional form f(R) shown in Figure \ref{fig:HAVERAH-1} and given by f(R) = $ \exp(\frac{-R}{\; \;R_o})$, with $R_o $ = (110+/-10) m for $\nu$ = 55 MHz and $\theta \leq 35^\circ$. $R_o$ represents the scale factor for radio lateral distribution pool, becoming larger as the zenith angle increases or as the observational frequency decreases.
These findings made it possible to formulate the simple relationship between \enu  and other array derived shower parameters   \enu =  $ 20 \left ( \frac{E_P}{10^{17}} \right )  \sin(\alpha)  \cos(\theta) \exp \left ( \frac{-R}{R_o(\nu,\theta)} \right ) \; \rm{in} \; \mu V \; m^{-1} MHz^{-1}$ within the limited range of distances R and zenith angle $\theta$. Such insights into the behaviour of the radio signature from EAS would not have been possible without direct access to array data. An immediate implication of the outcomes of this experiment was that radio pulse amplitudes were probably not determined or influenced to any important degree by any additional parameters, ruling out the possibility of radio being a potential tracer of primary cosmic ray mass.
 
\subsection{Exclusively radio triggering systems, long baseline stations and the radio frequency spectrum.}
Soon after confirming the radio phenomenon in 1965 \cite{POR65}, the UCD group became intrigued with the possibility of detecting very energetic EAS simply by radio methods alone \cite{FEG68A}. Ireland in the mid-1960s was remarkably radio-quiet and the opportunity to combine a number of experimental techniques in one rather ambitious plan proved compelling.Three independent observational stations were set up and operated.  A variety of VHF and UHF radio systems were assembled, with some ancillary shower detection equipment employed at two of the three stations, see Table \ref{Tb:Stats2}. 
\begin{table}
\caption{Long baseline stations - Sites, radio receiver operational frequencies (MHz), bandwidths (MHz) and ancillary detectors}
\label{Tb:Stats2}
\vspace{0.1in}
\begin{center}
\begin{tabular}{p{3.25cm}|p{0.5cm}|p{0.43cm}|p{2.0cm}} \hline
SITE & $\nu$ & \bw & Ancillary detectors\\[1ex] \hline \hline
UCD Campus (Urban, quiet at night)       & 12 45 45 70 520      & .05  4  4  20  20     &Cherenkov (sporadic) \& Particle  (continuous) \\ \hline
Glencullen (shielded valley)                     &  35 70 70                 & 10 15 15                  &Cherenkov (sporadic) \\ \hline 
Kilbride (quiet, remote, open site)            &70                             & 15                            &                                    \\ \hline
\end{tabular}
\end{center}
\end{table}
The stations were located at the corners of a triangle $44 \rm{km}^2$ in area. All antennae pointed towards geomagnetic west and very close to the horizon. A shower of $10^{18}$ eV, incident at $80^{\circ}$ zenith angle, was estimated to produce a fully coherent radiation pool at 75 Mhz of $10^2 \rm{km}^2$   but projected onto an ellipse on the ground of area approximately $60 \rm{km}^2$.   A novelty was to dispense with any particle detector or optical Cherenkov detectors in the local \emph{trigger} systems at each station and rely purely on locally generated radio triggers. This made it possible to operate in 24/7 mode with optical Cherenkov information available at two of the three stations on a sporadic basis, but with  very low duty cycle efficiency due to weather, moonlight etc. In addition, particle detector information was continuously recorded at one of the three stations. The RF outputs of all radio receivers and ancillary cosmic ray detectors were displayed on oscilloscope timebases which were continuously  photographed by motor driven cameras fitted with rolls of 35mm TriXpan film. 

\begin{figure}
\begin{center}
\includegraphics[width=.475\textwidth]{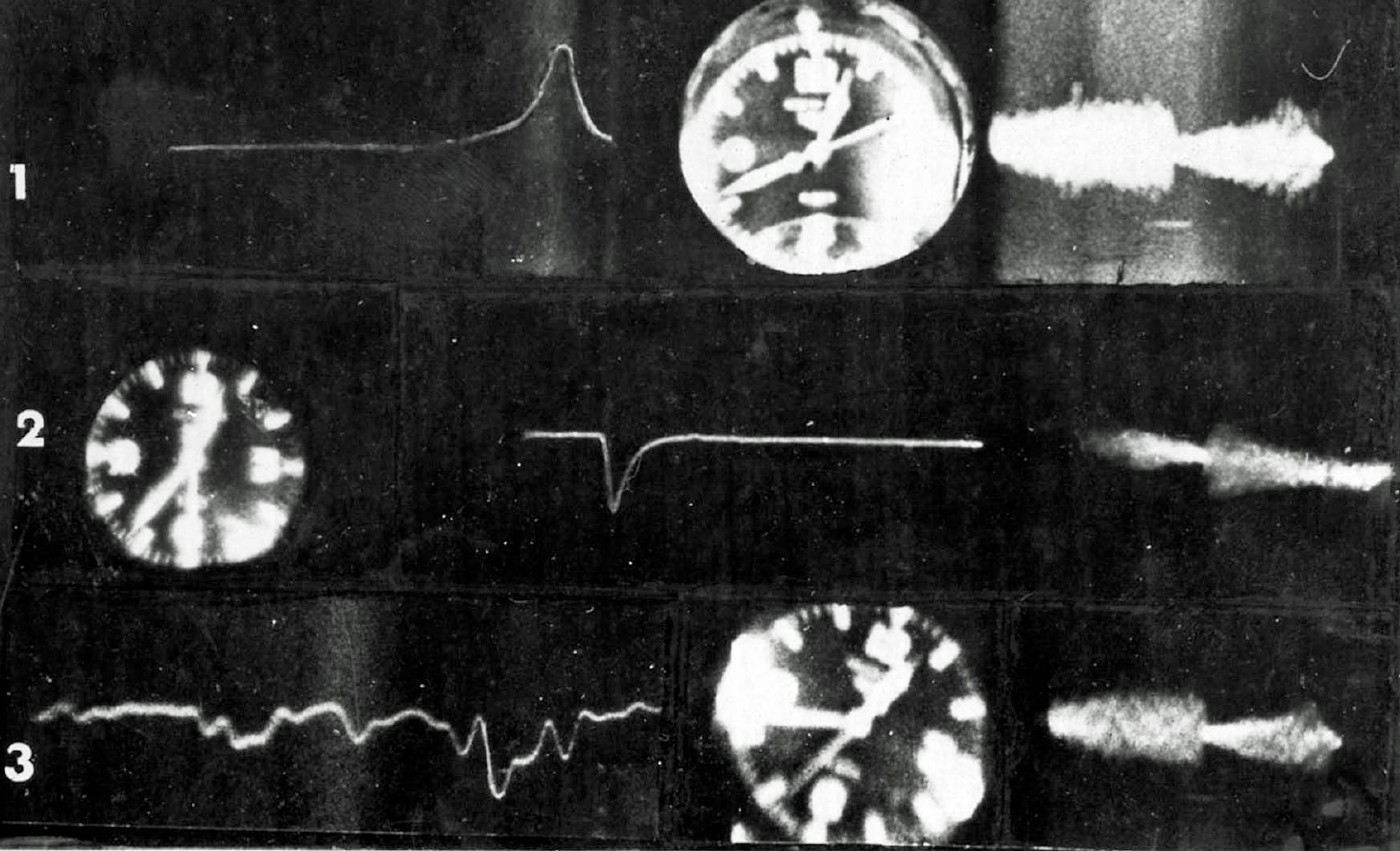}
\caption{A two-station long-distance coincident event. Waveform (1) Belfield 45 MHz \bwl pulse; Waveform (3) Belfield 70 MHz structured pulse; Waveform (2) Glencullen 70 MHz \bwl pulse. Note oscilloscope waveforms (1) and (3) run right-to-left, while waveforms (2) run left-to-right. The three modulated carrier waveforms at the extreme right represent the fine-timing broadcast signals}
\label{fig:TFC}
\end{center}
\end{figure}

Triggering at the UCD station was based on a prompt 2-fold coincidences between signals at 45MHz and for the Glencullen station  2-fold coincidence at 70 MHz. For the third station at Kilbride, the trigger was simply a pulse several times greater than ambient noise. Establishing simultaneity between events at different stations was not easy in the 1960Õs and had to be done offline, long after individual station events were registered. Initially 24-hour analog clocks at each station were photographed whenever a local trigger happened, by illuminating the clock faces with a lamp, turned on by a silicon controlled rectifier pulse derived from the master station trigger unit. These clocks were synchronised to an absolute accuracy of about +/- 1 second every 10 days, using broadcast low frequency time standards. They would then drift relative to one another, but in a measurable manner. However, close to millisecond fine-time synchronization was later achieved by additionally photographing the modulated carrier of the 200kHz BBC Light Program commercial radio transmission, on a separate oscilloscope at each station. Searches for multiple station coincident events were conducted laboriously by a human scanner searching through films for interesting events (+/- 2s) and printing off highly likely station candidates for more careful examination of the modulated 200kHz broadcast signal carrier timebase waveforms, which for coincident events would be identical within about the 50 ms resolution limit of the method (Figure \ref{fig:TFC}), for a typical 2-station coincidence.

The outcomes of the quest to find long-distance coincidences are summarised in Table \ref{Tb:Stats3}. The three individual baselines were Belfield-Glencullen (10km), Glencullen-Kilbride (12km) and Belfield-Kilbride (20km). Findings were to some extent compromised by logistical difficulties in operating such a complex experimental setup spread over three stations, two of which were quite remote from the UCD campus at Belfield, particularly the Kilbride station which suffered frequent AC power outrages.

\begin{table}
\begin{center}
\caption{Two-fold course timing (+/- 2s resolution) coincident event statistics, observed (Obs.) and randomly expected (Exp.). In the two rightrmost columns the fine timing (+/- 25 ms resolution) coincident event statistics, observed (FT-O) and randomly expected (FT-E).}
\label{Tb:Stats3}
\vspace{0.2cm}
\begin{tabular}{|c|c|c|c|c|c|} \hline
Baseline  & Livetime  &  Obs. &  Exp. & FT-O & FT-E \\ \hline \hline
10 km & 1024 hrs & 755 & 598 +/- 24 & 23 & 1.8 \\ 
12 km &   244 hrs &   48 &   52 +/- 7   &   0  & 0.05 \\
20 km &   160 hrs &  111 &  85 +/- 9    &   0  & 0.13 \\ \hline
\end{tabular}
\end{center}
\end{table}

The most important scientific conclusion was that no 3-fold coincidences were observed \cite{FEG68A}. However, it should be pointed out that the overlapping \fov between Glencullen and Kilbride stations was restricted to a maximum zenith angle of 84$^\circ$, due to mountainous terrain. Some positives and some negatives came out of this experiment which may be summarised as follows Ð
\emph{Positives:} (a) Triggering on EAS radio pulses alone at 45 MHz or 70 Mhz was feasible but interpretation of results was difficult. (b) At local stations, occasional radio triggers had accompanying signals either in the particle detector or the Cherenkov detectors, when it proved feasible to operate them. (c) Radio pulse shapes at 70 MHz and 45 Mhz were frequently very clean isolated single \bwl events, consistent with model predictions. (d) UHF radio emission at 520 MHz was detected for the first time and the integral pulse height spectrum of detected pulses suggested an incoherent emission mechanism \cite{FEG68B}.  
\emph{Negatives:} (a) No 3-fold coincident long-distance events were observed. (b) 2-fold long-distance coincident events were observed with pulse shapes consistent with EAS emission but there was no direct absolutely unambiguous verification that the coincidences were associated with EAS. (c) AGC rate-controlled radio trigger systems (45MHz at the UCD site) had, of necessity, continuously variable sensitivity thresholds making accurate pulse height spectra measurements meaningless or impossible for this observational frequency. (d) Cherenkov night-sky back-up systems had extremely low duty-cycles due to weather, moonlight and the necessity of human operation and offered little of direct benefit to sustained programs of radio observation. (e) In urban environments, man-made RF interference constituted a sporadic but real impediment to progress, implying that reliable measurements could only be made in radio quiet locations such as shaded woody valleys with low surrounding hills. Of course such terrain precluded observations at extremely large zenith angles. (f) The rate of detection of LZA radio triggers from showers of energy $ \ge 10^{18}$ eV with simultaneous pulses at 44MHz, 520 MHz and particle signature in the scintillator was only $\simeq 0.03\;  \rm{ hr^{-1}}$. With the termination of the long baseline radio work in late 1967, the UCD group decided to concentrate almost exclusively on \Gra development.  Radio investigation thereafter was restricted to operating in the exceptionally quiet UHF part of the spectrum and contributed \cite{FEG72} with other groups \cite{SPEN69}, \cite{CJ69}, \cite{C70} and \cite{GALL70} to establishment of the frequency spectrum of radio emission from showers.  Between 40MHz and 550 MHz, the average radio pulse energy normalised to 2 $\times 10^{15}$ eV was found to obey a frequency dependence of the form $\nu^{-2}$.

\section{\Ch detector systems 1967-1970} \label{lateC}  
\subsection{The Whipple Observatory 10m aperture reflector} 
\Ch detector sensitivity may be increased by lowering the energy threshold which depends on system parameters through the  relationship $   E_t \propto \frac{1}{ \epsilon D} \sqrt{  \frac{\Omega}{f \; \nu\ \delta \nu}  }$ where $\Omega$ is the system field-of-view, f the electronic bandwidth, $ \delta \nu$ the bandwidth at some central frequency $\nu$, $\epsilon$ the phototube quantum efficiency and D the detector's aperture. In 1967, Giovanni Fazio and  colleagues \cite{FAZ68} began construction of a new reflector that was commissioned the following year. Mainly exploiting the parameter D in the energy threshold equation, the new instrument, of unprecedented 10m aperture, was located at the Whipple Observatory in Arizona, at an altitude of 2300m. Design features incorporating a combination of huge aperture, fast electronics, high-efficiency S13 photomultiplier tubes, coupled with an exceptionally dark site, resulted in a lowering of the energy threshold by a factor of 10 over what was otherwise attainable anywhere, to between $2 - 4 \times 10^{11}$ eV. The angular acceptance $\Omega$ was approximately $5 \times 10^{-4}$ sr and the collection area of the telescope at $30^\circ$ was estimated as $2.5 \times 10^{8} \rm{cm^2}$.
\begin{figure}
\begin{center}
\includegraphics[width=.45\textwidth]{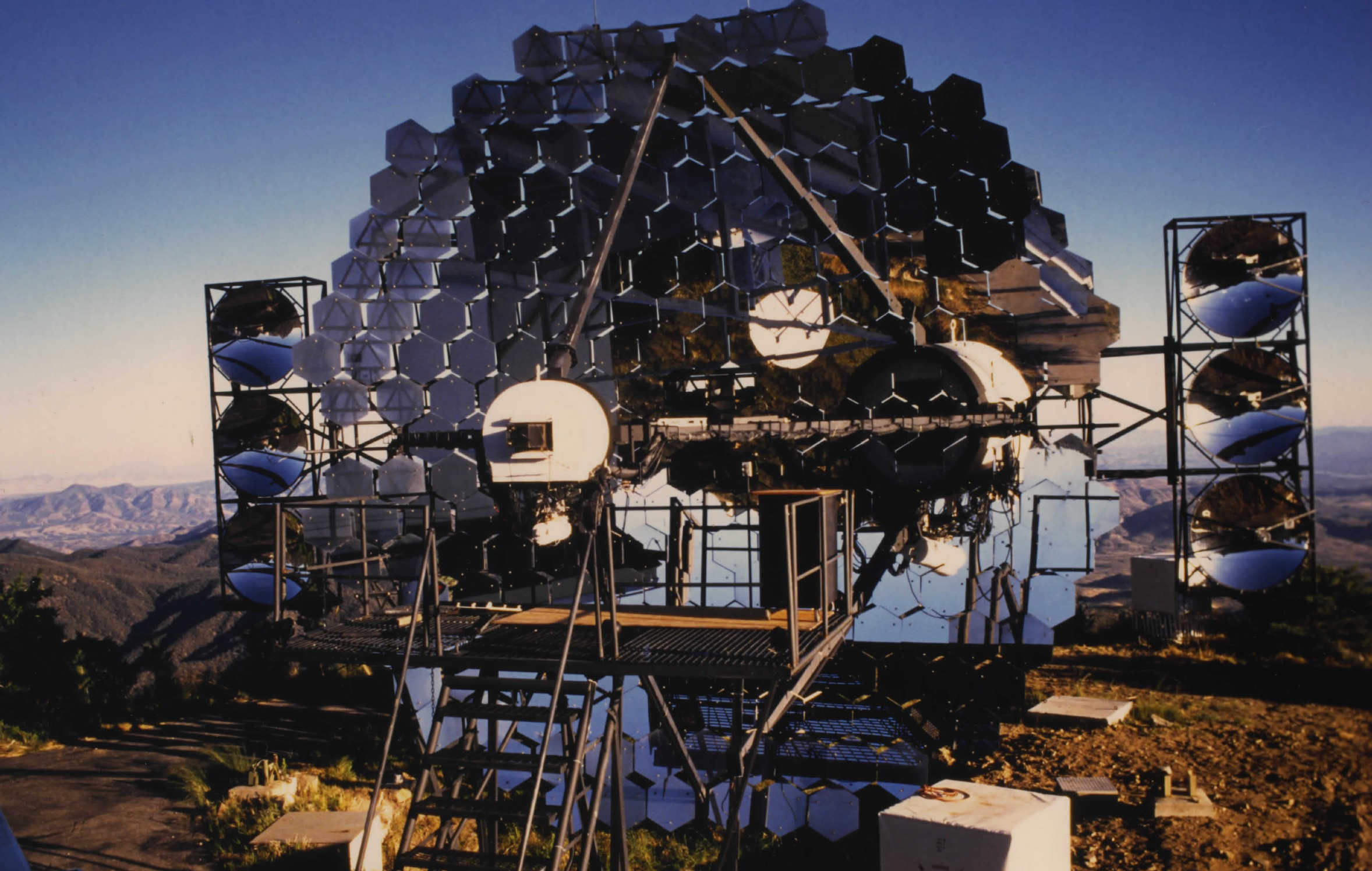}
\caption{The Whipple 10m reflector photographed during the 1990s with 6 small out-rider reflectors attached.}
\label{fig:WHIP10M}
\end{center}
\end{figure}
It is somewhat ironic that this exceptionally fine and wonderfully engineered instrument (Figure \ref{fig:WHIP10M}) gave service for almost two decades before the first successful and unambiguous detection of a TeV source was accomplished in 1988, as a result of the implementation of the Imaging Atmospheric Cerenkov Technique (IACT). The 10m instrument was absolutely central to the development of imaging and to the very long history of TeV \Gra and continues to play a role in the field, more than 40 years after commissioning.

\subsection{A 4-fold fast optical \Ch telescope}\label{SS:4-fold}
The phenomenon of \Ch radiation in the atmosphere was reviewed by Jelley in 1967 \cite{JVJ67} just at the time when the UCD group made the
strategic decision to greatly diminish research activity in radio detection of EAS. \Gra appeared to be a more promising bet and  the Harwell-UCD collaboration decided to build  on experiences gained during the first part of the 1960s (Section  \ref{SS:FirstPM}) and initiated a new phase of  collaboration. During the winter of 1966-67 a novel approach was experimented with at Glencullen, using a fast twin-mirror reflector system \cite{FEG68C}.
The new approach was strongly influenced by calculations and analytical modelling of the early stages of shower development.  High energy particles in the early generations of the electromagnetic cascade of photon-induced showers occur above 10km (from 20km down to 10km atmospheric height) and for energies $\geq 10^{10}$ eV will essentially be unscattered and will deliver light into a narrow annulus of 100 to 130m radius at sea level, with a time-spread of about 3ns - \emph{the fast directed light component}. This focussing effect arises from a somewhat fortuitous cancellation of the \Ch angle with the shower height factor. This fast component is superimposed on the rather slower arriving (20ns to 40 ns) pool of light from the gross shower, which illuminates a pool of radius about 250m. Since particle scattering is not entirely absent in the early stages of the cascade, some blurring of the ring into the larger pool will occur. In proton induced showers any fast component is expected to be less intense and less directed, since primary energy seeps out into the electron component over a considerably greater track trajectory. 

 \begin{figure}
\begin{center}
\includegraphics[width=.5\textwidth]{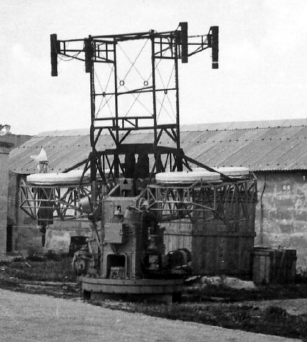}
\caption{4-fold fast optical \Ch telescope at Malta, 1969 \cite{JEN74}.}
\label{fig:MALTA-4}
\end{center}
\end{figure}

Here then was the potential to reduce threshold energy and improve angular resolution of \Ch detectors, ostensibly through narrowing the \fov over what had been conventional in the earlier generation prototype experiments. Speed was also an important exploitable parameter due to the very fine time-spread of the directed emission. Such considerations became prime drivers in the design of a completely new system, pioneered as a 3-mirror detector at Glencullen in 1967-68, moved to Harwell in the summer of 1968 where a fast-timing system for pulsar studies was added and then transported to a disused air-force base in Malta where full-scale observations began in February 1969. Configured as a 4-mirror reflector (Figure \ref{fig:MALTA-4}) consisting of four f/2 90cm mirrors, fast RCA 8575 phototubes, 2.5ns integration time amplifiers, a 3.5ns coincidence resolving time, the mirror mounting had a steerable Alt-Az drive with pointing accuracy of 4 arc minutes. The 4-fold design allowed running at high singles rates of $3 \times 10^5$ Hz with individually resolved photoelectrons. With a $1^{\circ}$ full \fov and energy threshold $E_t \sim 2 \times 10^{12}$, the 4-fold photon intensity at threshold was estimated as 30 photons m$^{-2}$ and with a flux sensitivity $ \sim$ 1 to 3 $\times10^{-11}$ photons \, cm$^{-2}$ s$^{-1}$. This extremely narrow \fov coupled with the very fast electronics greatly reduced background counting rates which bedevilled the prototype experiments that operated earlier in the decade, (Section \ref{S:earlyC}).
\subsection{Astrophysical performance of the fast 4-fold \Ch telescope}
The exciting discovery of pulsars led to an exclusive observational strategy focussing on these enigmatic objects. Analysis embraced searching for both aperiodic and periodic signals, following incorporation of fast accurate timing hardware. Barycentric correction methods, periodic search strategies and source ephemerides were all novel \emph{black arts} that had to be rapidly mastered. For the years 1969 and 1970 a comprehensive and rigorous program of drift-scan observations was undertaken on 10 pulsars. Disappointingly, despite the demanding observational program, none of the pulsars studied reflected consistent or significant positive effects, all being compatible with zero continuous flux of $\gamma$-rays. Individual  drift-scans were generally of duration 30 to 40 minutes,  symmetrically spanning up to (+/- 4.5$^{\circ}$) about the transit of  each source. Over the full set of observations aimed at finding \emph{directed} emission, individual  pulsars were within (+/- $0.75^{\circ})$ of transit for on-source exposure durations of anything between 50 and 750 minutes exposure. For \Gr astronomy, with its inherently limited duty cycle, these values corresponded with quite deep exposures on source, given the relatively unsophisticated technology of the era. Flux limits  $ \sim$ 0.5 to 2.5 $\times10^{-11}$ photons cm$^{-2}$s$^{-1}$ were established and eventually published in 1974, \cite{JEN74}. These negative results were compatible with those of the TATA group based on somewhat contemporaneous observations made at Ooty, India \cite{CHA71}.
 \begin{figure}
\begin{center}
\includegraphics[width=.4\textwidth]{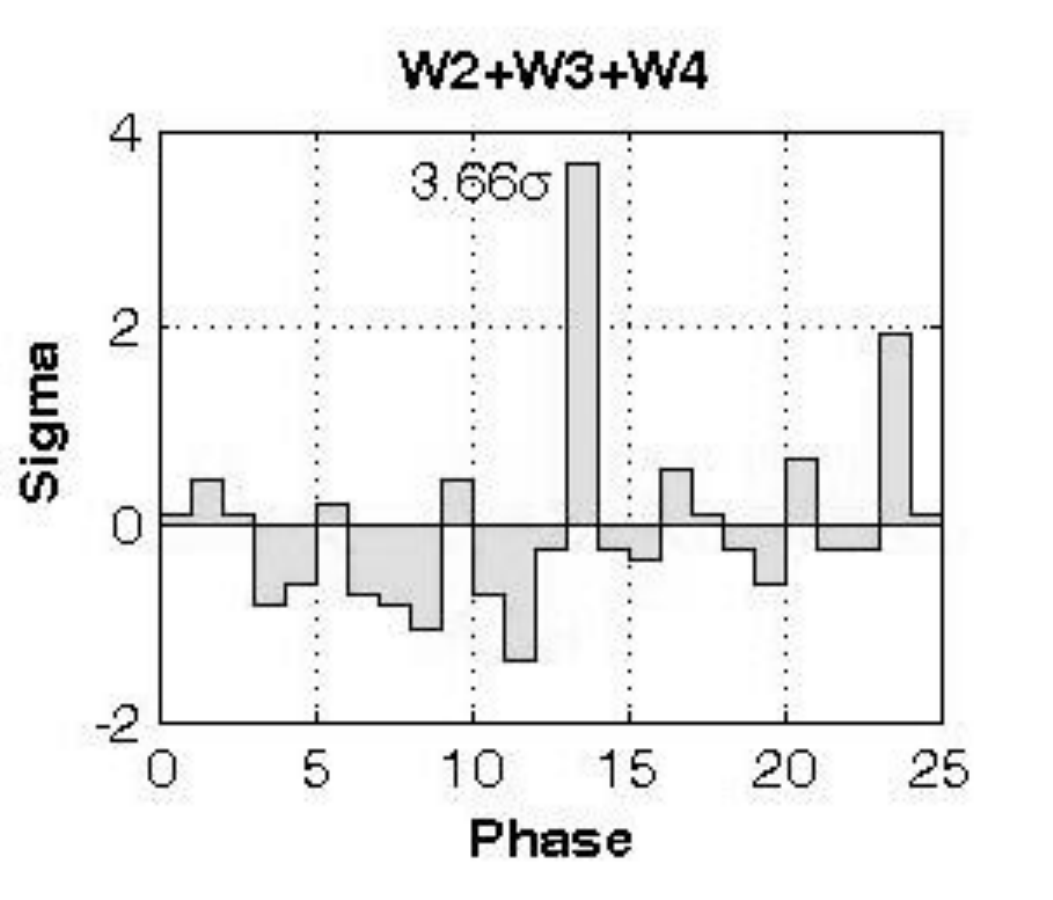}
\caption{Crab pulsar phaseplot, (25 bins, all ``phases'' multiplied by 25)}
\label{fig:CRAB33}
\end{center}
\end{figure}

\begin{figure}
\begin{center}
\includegraphics[width=.4\textwidth]{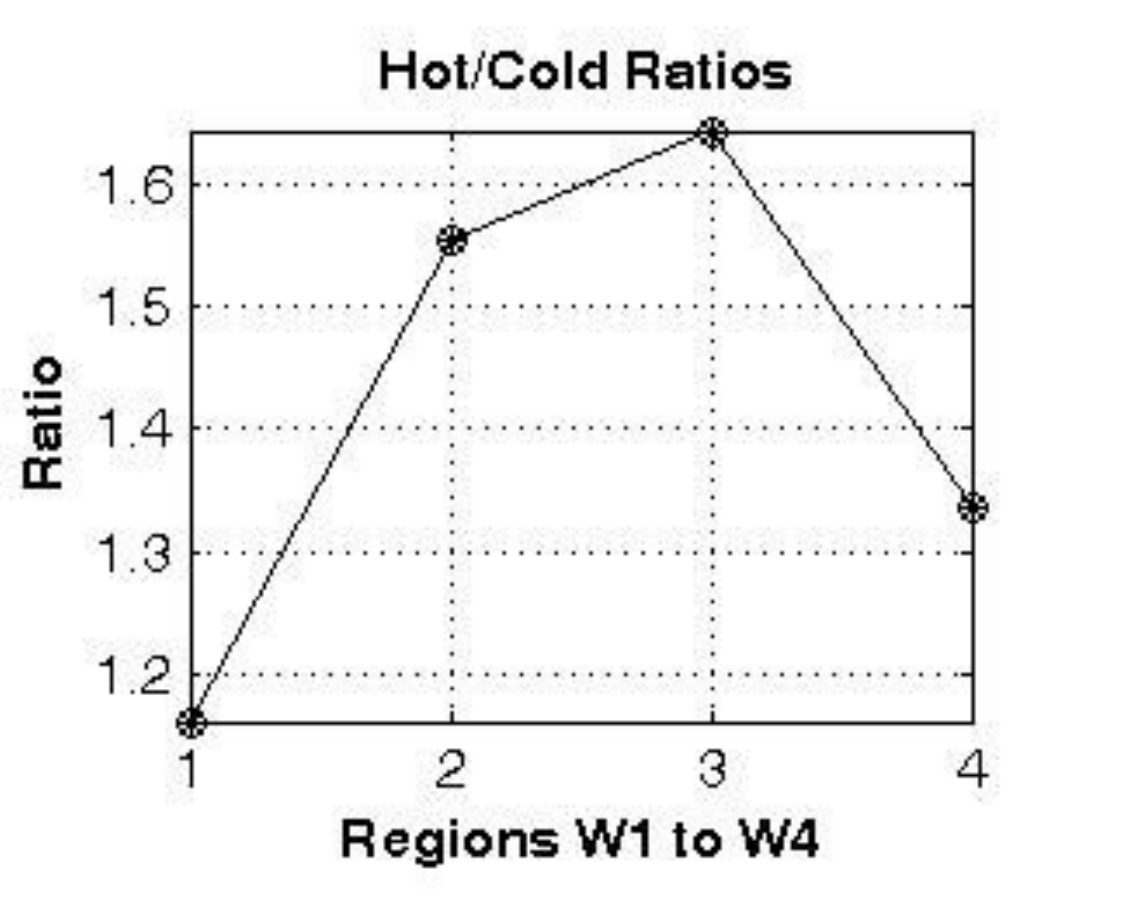}
\caption{\Ch Ratios of the \emph{mean} population of ``Hot'' phase bins (14 and 24) to ``Cold phase bins (all other 23 bins) as a function of the independent regions W1 W2, W3 and W4.}
\label{fig:HC-RATIO-2}
\end{center}
\end{figure}

One particular set of observations on the Crab pulsar ($P_0 = 33ms$) is worthy of highlighting, if only to exemplify the tedious nature of the drift-scan technique and the very meagre counting statistics by contemporary standards. A possible anomalous flux of pulsed \Grs  was observed in 12 hours of drift-scan data taken during January/February 1970. Data was analysed for periodicity by phase-linking over the entire sequence of observations and selecting regions in angular space around source transit. Region W1 corresponded to all the events falling within (+/- $0.83^{\circ}$); region W2 symmetrically spanned +/- ($0.83^{\circ}$ to $1.66^{\circ}$); region W3  symmetrically spanned +/- ($1.66^{\circ}$ to $2.49^{\circ}$) and region W4  spanned +/- ($2.49^{\circ}$ to something between  $4.5^{\circ}$  and $5.00^{\circ}$), symmetrically about transit. Arrival times of events in each of the four regions were barycentric corrected, divided by the Crab pulsar period and the individual event phases plotted as 25-bin phase histograms. Region W1 exhibited no statistically significant evidence for  pulsed emission but the three other regions all did. A combined events phaseplot for regions W2, W3 and W4 is shown in Figure(\ref{fig:CRAB33}). This plot exhibits two peaks separated by 13.2 ms, the observed separation of the main and inter-pulse for the Crab pulsar. The largest populated peak is 3.66${\sigma}$ above the mean, the second largest is 2.01 ${\sigma}$ above the mean. The overall random probability of the observation was conservatively calculated to be about $10^{-4}$. Assuming the observation corresponded to emission of genuine pulsed \Grs from the Crab pulsar, then what was being observed reflects a broad angular emission profile, while the original scientific objective of fast directed emission (predicated on using  a narrow \fov detector) was not realised. While the actual data has long since disappeared it was possible on the basis of the published results \cite{JEN74} to look retrospectively at the ratio of the \emph{mean} population of the two ``hot'' phase bins (14 and 24) to the \emph{mean} of the``cold'' phase  bins (all other 23 bins), for the four independent regions W1 to W4 as shown in Figure (\ref{fig:HC-RATIO-2}). The ratio rises from region W1 to maximise in region W3 and thereafter falls appreciably. This form of behaviour is consistent with what a genuine \Gr emission signature might be expected to exhibit. Such a broad profile is consistent with the expected  lateral distribution of \Ch light as predicted theoretically. By contemporary standards the event trigger rates were minuscule, the combined 12 hours Crab pulsar data corresponding to 1870 events, between 2 to 3 events per minute with this narrow \fov detector. However, the pulsar lightcurve as represented in phaseplot of Figure (\ref{fig:CRAB33}) is shifted by almost half a cycle from the absolute Crab pulsar lightcurve, an artifact that was possibly due to an inherent unresolved hardware timing issue within the electronics, but which was never resolved at the time. Interestingly, in Crab pulsar data taken in early 1973 by Grindlay and collaborators \cite{GR73} using the 10m reflector at Mt. Hopkins, a somewhat similar observation of possibly phase-shifted \Grs was made. 
\section{Looking in the rear-view mirror}
The decade between 1960 and 1970 was an exciting and promising one from the perspective of  development of radio and optical \Ch techniques. However, it was also frustrating in terms of limited financial backing, unsophisticated technological infrastructure and limited computational facilities.  Radio promised much, but following an initial spurt of activity it became clear that emission was much more highly beamed than originally suspected and the prospect of stand-alone radio receivers for shower detection ultimately proved unfeasible, so that by 1975 interest had well and truly waned. The  scientific return, in  proportion to the efforts of those participating, was modest \cite{Allan71}. A quarter of a century later the emergence of a new astrophysics challenge came chasing after the old, if unproven radio technique. This entirely new challenge, prompted by the RADHEP 2000 (Radio Detection of High Energy Particles) International Workshop in Los Angeles, revived interest in the technique as a tool for Neutrino Astrophysics and a resurgence of interest in radio was flagged \cite{FG2003}. By the turn of the new millennium radio and communications technology had become very sophisticated and signal processing and digital filtering were inexpensive  and relatively straight forward to implement. The radio technique is now well and truly prospering, as this ARENA 2010 workshop testifies.

In contrast, the optical \Ch work never absolutely terminated  but rather went into slow decline through much of the 1970's, only to be resuscitated on the basis of the very important paper of Weekes and Turver \cite{WT77}. Here the value of shower simulations was patently evident, as it was shown that the form of the \emph{average} lateral distribution of light from \Grs was different from that of proton induced showers. Proton induced showers are deficient in light content at 100GeV in comparison with $\gamma$-rays. Here at last was a possible basis for  \emph{background discrimination}, whereby unwanted proton background showers might be eliminated using both hardware and software techniques. The suggestions here proposed the use of twin reflectors operating in stereoscopic mode, each configured with 19 or 37 phototube arrays, located at the prime foci, acting as 2-D image formation segmented photon collectors. As stated earlier however, it was almost a further decade before the imaging technique became established beyond reasonable doubt. Could progress in TeV \Grs have been more rapid? Probably not. The long history of upper limit experiments, coupled with both the poor flux sensitivities of the early detectors and the absence of sophisticated technology, may all have mitigated against serious agency funding. This situation only gradually changed in the mid-1980's. Had not the Whipple collaborators benefitted during that decade from unlimited access to the enduring 10m reflector, while at the same time slowly establishing an ever more  stable imaging camera and simultaneously accessing increasingly more sophisticated sets of simulations (less fluctuation prone) to underpin image parameterization, then progress in the field of TeV astrophysics might have been even slower than history testifies! 

However, without doubt, the ultimate revival of both the radio and optical fields was firmly rooted in the foundational and innovative experiments of the 1950's and 1960's. At the 70th birthday symposium for Neil Porter in 2000, the accomplishments of those earlier times were perceptively summarised  by Vladimir Vassiliev in the following words (unpublished) -
`` The majority of ideas behind the Atmospheric \Ch technique were formulated in the fifties and sixties, including those of imaging and stereoscopy, procedures for `stabilizing' the spatial variations of the apparent brightness of the night sky, techniques for monitoring sky transparency by recording light from a reference star, as well as methods of night sky background suppression by utilising multiple coincident triggers. The founders of this technology have not only provided the basis for the development of the next generation of ground-based observatories, such as VERITAS, they also have given to us exceptional examples of depth and clarity of thinking''. Remarkably, many of the pioneers of the \Ch technique also contributed significantly to the development of the radio technique. It is to be desired that the radio technique (with acoustic alternatives), will aspire to similar astrophysical successes as those so far accomplished by the optical \Ch technique.




\bibliographystyle{elsarticle-num}
\bibliography{<your-bib-database>}






\end{document}